\documentclass[%
 reprint,
 amsmath,amssymb,
 aps,
 prl
]{revtex4-1}

\usepackage{graphicx}
\usepackage{dcolumn}
\usepackage{bm}
\usepackage[mathlines, columnwise]{lineno}

\usepackage{braket}
\usepackage{color}

\newcounter{lastnote}

\begin{document}
\newcommand{\beginsupplement}{%
        \setcounter{table}{0}
        \renewcommand{\thetable}{S\arabic{table}}%
        \setcounter{figure}{0}
        \renewcommand{\thefigure}{S\arabic{figure}}%
        \setcounter{equation}{0}
        \renewcommand{\theequation}{S\arabic{equation}}%
     }
\preprint{APS/123-QED}

\title{Two-mode Dicke model from non-degenerate polarization modes}

\author{Andrea Morales}
\affiliation{Institute for Quantum Electronics, ETH Zurich, CH-8093 Zurich, Switzerland}
\author{Davide Dreon}
\affiliation{Institute for Quantum Electronics, ETH Zurich, CH-8093 Zurich, Switzerland}
\author{Xiangliang Li}
\affiliation{Institute for Quantum Electronics, ETH Zurich, CH-8093 Zurich, Switzerland}
\author{Alexander Baumg\"artner}
\affiliation{Institute for Quantum Electronics, ETH Zurich, CH-8093 Zurich, Switzerland}
\author{Philip Zupancic}
\affiliation{Institute for Quantum Electronics, ETH Zurich, CH-8093 Zurich, Switzerland}
\author{Tobias Donner}
\email{donner@phys.ethz.ch}
\affiliation{Institute for Quantum Electronics, ETH Zurich, CH-8093 Zurich, Switzerland}
\author{Tilman Esslinger}
\affiliation{Institute for Quantum Electronics, ETH Zurich, CH-8093 Zurich, Switzerland}

\begin{abstract}
We realize a non-degenerate two-mode Dicke model with competing interactions in a Bose-Einstein condensate (BEC) coupled to two orthogonal polarization modes of a single optical cavity. The BEC is coupled to the cavity modes via the scalar and vectorial part of the atomic polarizability. We can independently change these couplings and determine their effect on a self-organization phase transition. Measuring the phases of the system, we characterize a crossover from a single-mode to a two-mode Dicke model. This work provides perspectives for the realization of coupled phases of spin and density.
\end{abstract}

\maketitle
The Dicke model captures the coupling between a single electromagnetic field mode and an ensemble of two-level atoms \cite{Dicke1954}. This paradigmatic model is central to many developments in quantum optics \cite{Garraway2011,FriskKockum2019}. It also makes a connection to concepts usually studied in the context of condensed matter physics \cite{Diehl2008,Sachdev2011}, since it predicts, for strong enough coupling, a phase transition from a normal to a superradiant state even at zero temperature \cite{Hepp1973,Wang1973,Larson2017}. Theoretical investigations of the Dicke model and its variants \cite{Tolkunov2007, Gopalakrishnan2011, Strack2011a, Mivehvar2017, Quezada2017a, Liu2019} have given insights into critical behavior of open quantum many-body systems \cite{Diehl2008,  DallaTorre2013, Chitra2015, Soriente2018}, chaos \cite{Emary2003}, enhanced symmetries \cite{Baksic2014, Fan2014, Moodie2018}, and multi-partite entanglement \cite{Garraway2011,Brandes2004,Milburn2002}.

A few years ago, the Dicke model was realized experimentally in a driven-dissipative system coupling the external degree of freedom of a Bose-Einstein condensate (BEC) to an optical cavity via the atomic scalar polarizability \cite{Baumann2010}. Since then, experiments have strived to extend their possibilities in order to realize interesting variants of the single-mode Dicke model. In one approach, multi-mode Dicke models are engineered by coupling the atomic density to multiple cavity modes via the scalar atomic polarizability \cite{Leonard2017, Morales2017a, Kollar2017}. In a second approach, single-mode Dicke spin-models are realized, exploiting the vectorial atomic polarizability \cite{ Zhiqiang2016, Landini2018, Kroeze2018a, Zhang2018a}.

In this letter we combine these two concepts in order to achieve competing interactions between density and spin. We realize a non-degenerate two-mode Dicke model by coupling a BEC to the two fundamental polarization modes of a single cavity both via the scalar and vectorial polarizabilities where we can tune their interaction strengths independently. From a quantum simulation perspective, our system delivers the basic ingredients for the realization of condensed matter models with competing spin and density order parameters \cite{Cheong2007, Demler2004}, where, for example, open questions concern the scaling properties close to a multicritical point \cite{Narayan2017}.

\begin{figure}[t]
	\centering
		\includegraphics[width=85mm]{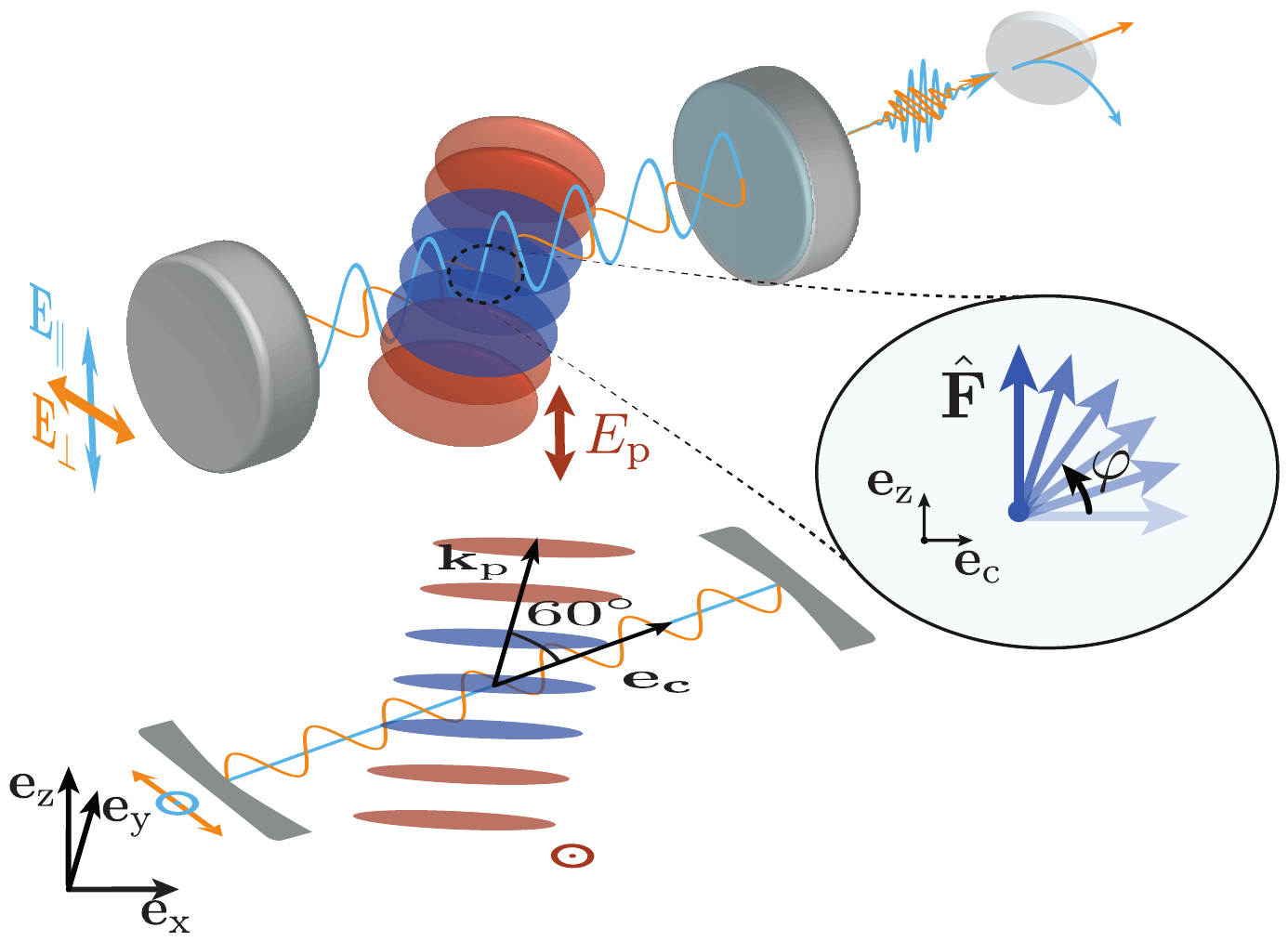}
	\caption{ \textbf{Engineering tunable vectorial and scalar couplings in a quantum gas coupled to an optical cavity.} A Bose-Einstein condensate (BEC) of $^{87}$Rb atoms (in blue) is optically trapped at the center of an optical cavity with wave-vector ${\bf k}_\textrm{c}=k_\textrm{c}{\bf e_\textrm{c}}$. The BEC is illuminated by a red-detuned optical lattice beam with wave-vector ${\bf k}_\textrm{p}=k_\textrm{p}{\bf e_\textrm{y}}$ and with wavelength $\lambda_\textrm{p}=785.5\,\;$nm, which we refer to as transverse pump. Its frequency is close to the resonance frequencies of two birefringent modes ($\parallel$,$\perp$) of the cavity (in light blue and orange) which are separated in frequency by 3.89(1)\,\;MHz.The pump and the cavity are tilted at $60^\circ$ and $k_\textrm{p}=k_\textrm{c}$. The polarization of the transverse pump electric field $E_\textrm{p}$ is linear and oriented along $z$, parallel to the electric field $E_\parallel$ of the the vertically polarized cavity mode and orthogonal to the electric field $E_\perp$ of the horizontally polarized cavity mode. The atomic pseudo-spin $\hat{\bf F}$ can be oriented in the ${\bf e}_\textrm{z}-{\bf e}_\textrm{c}$ plane by a bias magnetic field as described by the angle $\varphi$. Due to the finite reflectivity of the cavity mirrors, intra-cavity photons in the modes $\perp$ and $\parallel$ leak from the cavity in free space. Using a polarizing beam splitter placed on the axis of the cavity and two single-photon counting modules, we can detect photons in the cavity modes in real-time. }
	\label{fig:scheme}
\end{figure}

We conduct self-organization experiments of a spin polarized $^{87}$Rb BEC to two TEM$_{00}$  modes of an ultra-high finesse optical cavity with orthogonal polarizations, which we label as $\perp$ and $\parallel$. The atomic cloud couples to the $\perp$ mode via the vectorial atom-light coupling, and to the $\parallel$ mode, independently, via the scalar atom-light coupling. The atom-light interaction for each atom in the BEC is captured by an atomic dipole operator \cite{Kien2013} which can radiate in either of the cavity modes as described by the interaction energy (see supplementary material),
\begin{align}
\hat{H}_\textrm{int}=-\alpha_\textrm{s}\hat{\bf E}^\dagger\cdot\hat{\bf E}+i\alpha_\textrm{v}\Bigl(\hat{\bf E}^\dagger\wedge\hat{\bf E}\Bigl)\cdot\frac{{\bf \hat{F}}}{2F}
\end{align}
where $\alpha_\textrm{s}, \alpha_\textrm{v}$ are the scalar and vectorial atomic polarizabilities, $\hat{\bf F}$ is the atomic pseudo-spin vector operator and $\hat{\bf E}$ the total electric field operator. The vectorial part of the interaction can be controlled independently from the scalar part via the atomic spin vector $\hat{\bf F}$.\\
Our experimental setup is sketched in Fig.~\ref{fig:scheme}. The BEC is illuminated with an off-resonant standing wave laser beam of angular frequency $\omega_\textrm{p}$. This beam is referred to as \textit{transverse pump} and is angled at $60^\circ$ with respect to the cavity and polarized along ${\bf e}_\textrm{z}$. The $\parallel$ and $\perp$ modes have polarizations parallel and orthogonal to the transverse pump polarization, respectively. They are separated by a frequency difference $\delta=\omega_\parallel-\omega_\perp=2\pi\times3.89(1)\,\;$MHz,  due to birefringence. This frequency scale is large compared to the line-width of the cavity $\kappa/(2\pi)=147(4)\,$kHz.

The total hamiltonian describing the BEC-cavity system is (see supplementary material)
\begin{align}
\hat{H}&=-\hbar\Delta_\parallel\hat{a}^\dagger\hat{a}-\hbar\Delta_\perp\hat{b}^\dagger\hat{b}+\hbar\omega_\textrm{rec}\hat{c}^\dagger\hat{c}\notag\\
&+\alpha_\textrm{s}\frac{E_\textrm{p}E_0}{2\sqrt{2}}\Bigl(\hat{a}^\dagger+\hat{a}\Bigl)\Bigl(\hat{c}^\dagger\hat{c}_0+\textrm{h.c.}\Bigl)\notag\\
&+i\alpha_\textrm{v}\Bigl[\frac{E_\textrm{p}E_0}{2\sqrt{2}}\Bigl(\hat{b}^\dagger-\hat{b}\Bigl)\Bigl(\hat{c}^\dagger\hat{c}_0+\textrm{h.c.}\Bigl)\notag\\
&+\frac{E^2_0}{2}\Bigl(\hat{a}^\dagger\hat{b}-\hat{b}^\dagger\hat{a}\Bigl)\,\hat{c}^\dagger_0\hat{c}_0\Bigl]\,\frac{m_\textrm{F}}{2F}\,\textrm{cos}\,\varphi.
\label{spinHamiltonian}
\end{align}
The annihilation (creation) operators of an atom in the BEC are given by $\hat{c}_0$ ($\hat{c}^\dagger_0$). The annihilation (creation) operators of an atom in the momentum superposition resulting from scattering photons between pump and cavity are given by $\hat{c}$ ($\hat{c}^\dagger$). The operators $\hat{a}, \hat{b}$ ($\hat{a}^\dagger,\hat{b}^\dagger$) are the annihilation (creation) operators of photons in the cavity modes $\parallel$ and $\perp$ (with electric field amplitude $E_{\parallel}=E_{\perp}=E_0$ for a single intra-cavity photon). $E_\textrm{p}$ is the electric field amplitude of the transverse pump. The single photon recoil frequency is $\omega_\textrm{rec}=2\pi\times3.77\,$kHz. The quantities $\Delta_{\perp/\parallel}=\omega_\textrm{p}-\omega_{\perp/\parallel}-NU_0/(2\hbar)$ are the detunings of the transverse pump from the dispersively shifted cavity resonances of the modes $\perp$ and $\parallel$. The total atom number is $N$ with number operator $\hat{N}=\hat{c}_0^\dagger\hat{c}_0+\hat{c}^\dagger\hat{c}$, $\hbar$ is the reduced Planck constant, and $m_\textrm{F}=-F,..,F$ labels the magnetic sub-levels in the pseudo-spin manifold $F$.

The second line of the hamiltonian describes the scalar part of the coupling. Here, the BEC couples to the real (or in-phase) quadrature $\bigl(\hat{a}+\hat{a}^\dagger\bigl)$ of the vertically polarized mode of the cavity. When $\Delta_\parallel<0$ and for strong enough coupling, this term can drive the system into a self-organization phase (SO$_\parallel$) \cite{Baumann2010}. In fact, when a density fluctuation occurs in the BEC, a weak light field is scattered by the atoms into the cavity and builds up an intra-cavity field. At the position of the density fluctuation, the phase shift of the intra-cavity field relatively to the scattered field is zero when $\Delta_{\parallel}<0$ such that the resulting potential enhances the density fluctuation. For $\Delta_{\parallel}>0$ the phase shift is $\pi$ and density fluctuations are suppressed. Therefore, self-organization can only happen for $\Delta_{\parallel}<0$.

In between the squared brackets, two terms describe the vectorial part of the coupling. Both involve a cavity electric field of the form $i\Bigl(\hat{b}^\dagger-\hat{b}\Bigl)$ and therefore describe coupling to the imaginary (or out-of-phase) quadrature of the $\perp$ polarized cavity mode. When $\Delta_\perp<0$, the first term can also drive the system into a self-organization phase (SO$_\perp$). The second term describes scattering from the $\parallel$ mode to the $\perp$ mode and viceversa, with rate $\alpha_\textrm{v}E^2_0/2(m_\textrm{F}/(2F)\textrm{cos}\,\varphi)$. Differently from the scalar part of the coupling, the vectorial part can be controlled via the angle $\varphi$ between $\hat{\bf F}$ and the cavity axis ${\bf e}_\textrm{c}$ and can therefore be tuned independently.

The experiment starts with a BEC of $3.5(3)\times10^{5}$ atoms of $^{87}$Rb which is positioned at the center of the fundamental mode of the optical cavity (see Fig.~\ref{fig:scheme}). The transverse pump lattice wavelength is set to $\lambda=785.5\,\;$nm, where the ratio $\alpha_\textrm{v}/\alpha_\textrm{s}=1.085$ \cite{Kien2013}. The BEC is prepared in the atomic spin state $|F=1,m_F=-1\rangle$ and spin changing processes are suppressed by a large Zeeman shift (see supplementary material).
\begin{figure}[t]
	\centering
		\includegraphics[width = \columnwidth]{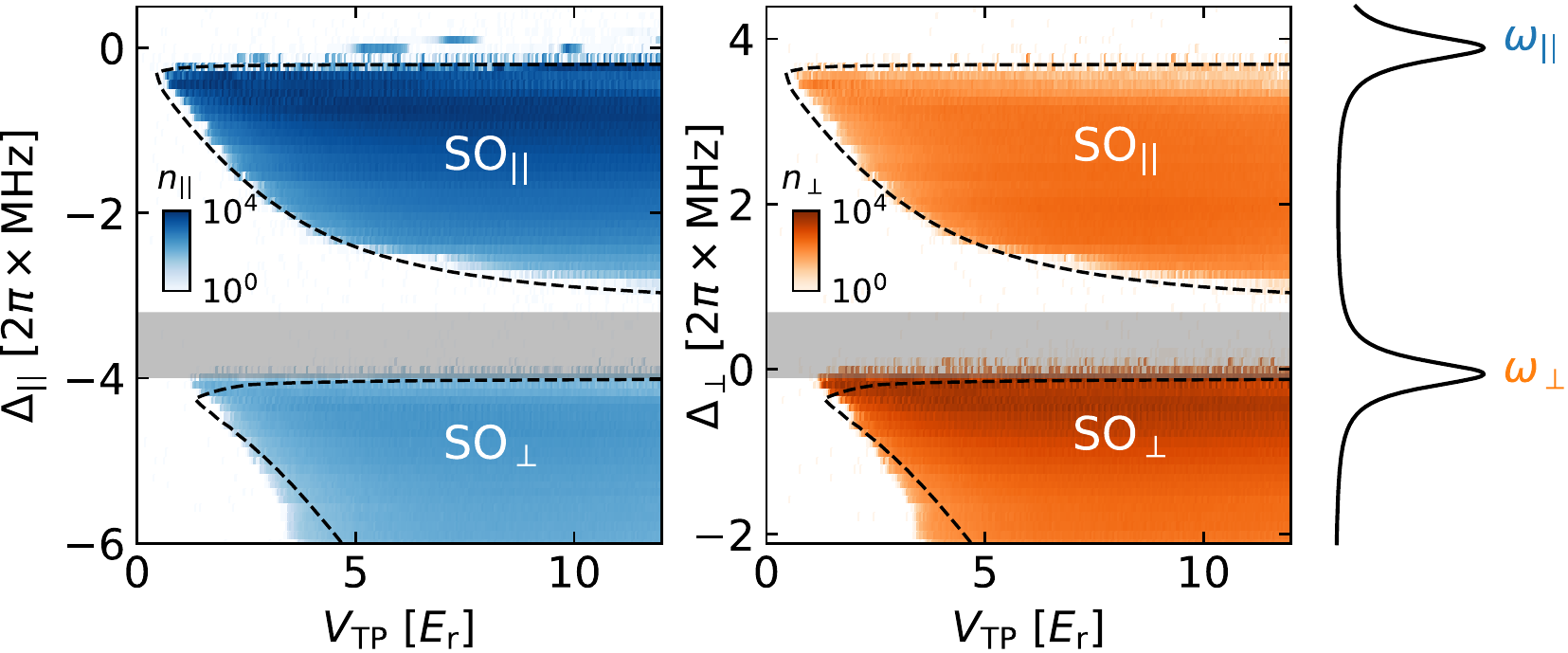}
	\caption{ \textbf{Phase diagram of scalar and vectorial self-organization in a birefringent cavity.} At $\varphi=0 (4)^\circ$, the calibrated intra-cavity photon numbers in the $\parallel$ (left panel) and $\perp$ (right panel) mode are recorded for different cavity detunings during ramps of $50\,$ms of the transverse pump (TP) lattice depth from $0$ to $12\,\hbar\omega_\textrm{rec}$. Experimentally we change the cavity detuning in steps of $100\,$kHz. The intra-cavity and transverse pump lattice depths are calibrated performing Raman-Nath diffraction on the BEC, and the dispersive shift is measured independently (see supplementary material). The black dashed lines are the predictions for the phase boundaries of the self-organization phases resulting from our theoretical model. The grey area indicates the detunings at which self-organization is suppressed by the simultaneous coupling to both polarization modes. On the right side, the solid black line indicates the cavity resonances for the birefringent modes. }
	\label{fig:PDFieldAlongCavityAxis}
\end{figure}
\begin{figure}[b]
	\centering
		\includegraphics[width=\columnwidth]{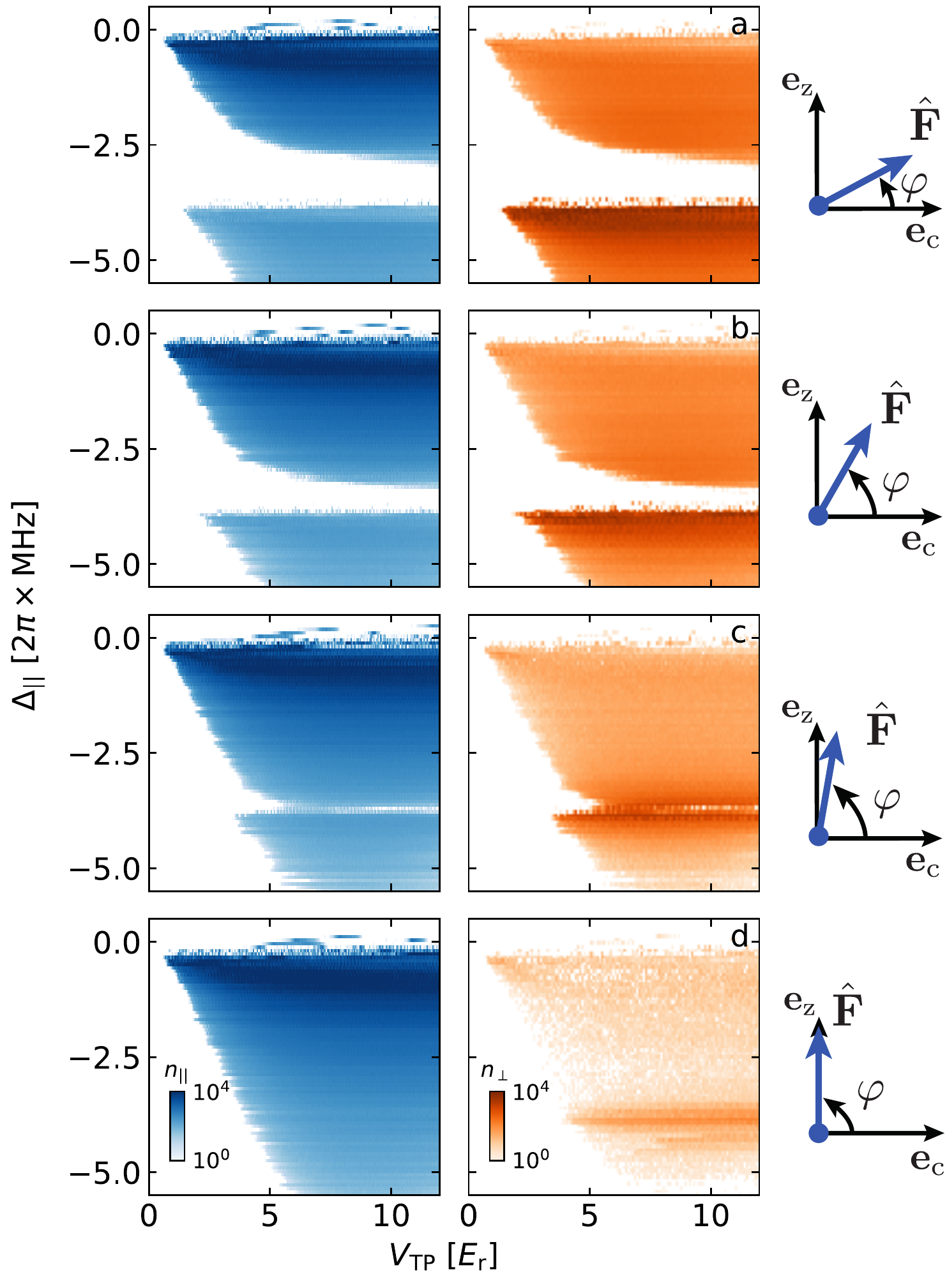}
	\caption{ \textbf{Tuning the vectorial atom-light coupling.}  {\bf a-d,} Starting from a situation where the magnetic field $B$ is oriented at $\varphi=30(4)^\circ$ in the ${\bf e}_\textrm{z}-{\bf e}_\textrm{c}$ plane and scalar and vectorial components of the polarizability are similarly strong ({\bf a}), we orient the atomic spin $\hat{\bf F}$ at the angles $\varphi=\,60(4)^\circ,\,80(4)^\circ,\,90(2)^\circ$ ({\bf b, c, d}). When the magnetic field ${\bf B}$ is parallel to the $z-$axis ({\bf d}), the vectorial coupling vanishes and scattering in the $\perp$ mode is suppressed. We attribute the residual feature in the $\perp$ mode to imperfect alignment of the pump polarization with the atomic spin.}
	\label{fig:OrientingB}
\end{figure}
We control the direction of the atomic spin in three dimensions by applying a magnetic offset field generated with four pairs of coils. The magnetic field direction and amplitude are calibrated performing radio-frequency spectroscopy on the BEC (see supplementary material). We monitor the intra-cavity photon number via the light field leaking out of the cavity.
The polarization of the intra-cavity light field is analyzed by placing a polarizing beam splitter at the cavity output that directs $\perp$ and $\parallel$ photons to two independent single photon counting modules.

In a first experiment, we orient the magnetic field along the cavity axis such that both the scalar and vectorial atom-light couplings are maximal (see Eq.~\ref{spinHamiltonian}). For fixed detunings $\Delta_\perp$ and $\Delta_\parallel=\Delta_\perp-\delta$ we ramp up in $50\,$ms the transverse pump lattice depth ($V_\textrm{TP}$) from zero to $12\,\hbar\omega_\textrm{rec}$, while simultaneously recording the output of the cavity in real-time. The measured intra-cavity photon numbers are reported in Fig.~\ref{fig:PDFieldAlongCavityAxis} for different cavity detunings. The two panels show the photons detected on the $\parallel$ and $\perp$ polarization modes of the cavity. In each panel, two phases are immediately visible, corresponding to the phases SO$_\parallel$ and SO$_\perp$. In the SO$_\parallel$ phase, we detect photons in the $\perp$ mode even when $\Delta_\perp>0$. For smaller values of $\Delta_\perp>0$, the critical pump lattice depth diverges and in a finite region of cavity detunings self-organization is forbidden.

Our results can be understood by mapping the hamiltonian in Eq.~\ref{spinHamiltonian} on a two-mode Dicke model \cite{Dmitry2007,Fan2014, Moodie2018}. We define $\hat{\beta}=i\hat{b}$ and then perform a rotation in the space of the photonic operators according to the unitary transformation
\begin{align}
\hat{a}&=\hat{t}\,\textrm{cos}\,\theta+\hat{d}\,\textrm{sin}\,\theta\notag\\
\hat{\beta}&=-\hat{t}\,\textrm{sin}\,\theta+\hat{d}\,\textrm{cos}\,\theta.
\label{UnitaryT}
\end{align}
Choosing the angle $\theta=\frac{1}{2}\textrm{tan}^{-1}\,\Bigl(\frac{\lambda_{\perp\parallel}N_0\,\textrm{cos}\,\varphi}{-\hbar\,(\delta/2)}\Bigl)$, where $\lambda_{\perp\parallel}=\frac{\alpha_\textrm{v}}{2}E^2_0\frac{m_\textrm{F}}{2F}$, the resulting two-mode Dicke hamiltonian (see supplementary material) reads as
\begin{align}
\hat{H}=&-\hbar\Delta_{t}\hat{t}^\dagger\hat{t}-\hbar\Delta_{d}\hat{d}^\dagger\hat{d}+\hbar\omega_\textrm{rec} \hat{J}_\textrm{z}\notag\\
&+ \frac{\lambda_{t}}{\sqrt{N}}\Bigl(\hat{t}^\dagger+\hat{t}\Bigl)\hat{J}_\textrm{x}+\frac{\lambda_{d}}{\sqrt{N}}\Bigl(\hat{d}^\dagger+\hat{d}\Bigl)\hat{J}_\textrm{x},
\label{TwoModeDicke}
\end{align}
where we have introduced the quantities
\begin{align}
\Delta_{t}&=\Delta_\parallel\,\textrm{cos}^2\theta+\Delta_\perp\,\textrm{sin}^2\theta-\frac{\lambda_{\perp\parallel}}{\hbar}\,N_0\,\textrm{sin}\,(2\theta)\,\textrm{cos}\,\varphi\notag\\
\Delta_{d}&=\Delta_\parallel\,\textrm{sin}^2\theta+\Delta_\perp\,\textrm{cos}^2\theta+\frac{\lambda_{\perp\parallel}}{\hbar}\,N_0\,\textrm{sin}\,(2\theta)\,\textrm{cos}\,\varphi\notag\\
\lambda_{t}&=\lambda_\textrm{s}\,\textrm{cos}\,\theta+\lambda_\textrm{v}\,\textrm{sin}\,\theta\,\textrm{cos}\,\varphi\notag\\
\lambda_{d}&=\lambda_\textrm{s}\,\textrm{sin}\,\theta-\lambda_\textrm{v}\,\textrm{cos}\,\theta\,\textrm{cos}\,\varphi,\notag
\end{align}
and the pseudo-spin operators $\hat{J}_\textrm{z}=\hat{c}^\dagger\hat{c}$ and $\hat{J}_\textrm{x}=\bigl(\hat{c}^\dagger\hat{c}_0+h.c.\bigl)$. We have defined $\lambda_\textrm{s}=\frac{E_\textrm{p}E_{0}\alpha_\textrm{s}\sqrt{N}}{2\sqrt{2}}$ and $\lambda_\textrm{v}=\frac{E_\textrm{p}E_{0}\alpha_\textrm{v}\sqrt{N}}{2\sqrt{2}}\bigl(\frac{m_\textrm{F}}{2F}\bigl)$. This hamiltonian describes the coupling of a macroscopic pseudo-spin $\hat{J}$ to two modes of the electromagnetic field described by the operators $\hat{t}$ and $\hat{d}$.

From this hamiltonian one can calculate the equation of motion of the system in a mean field theory where the operators are substituted by their expectation values. A self-organized phase corresponds to a non-zero steady state value of the average value of $\hat{J}_\textrm{x}$, or equivalently, to a non-zero average photon level in either cavity mode. Since the photons scattered into each cavity mode provide an optical potential that evolves on a time scale much faster than the atomic dynamic, we can perform an adiabatic elimination of the cavity fields and obtain an effective steady state solution for $J_\textrm{x}=\langle\hat{J}_\textrm{x}\rangle$ (see supplementary material) \cite{Ritsch2013}. Following this procedure, we obtain the equation
\begin{equation}
J_\textrm{x}=\pm\frac{N}{2}\sqrt{1-\frac{\omega^2_\textrm{rec}}{\eta^2}},
\end{equation}
where the phase boundary of the self-organization phase transition is fixed by the condition
\begin{equation}
r\equiv1-\Bigl(\frac{\lambda_t}{\lambda^{\textrm{crit}}_{t}}\Bigl)^2-\Bigl(\frac{\lambda_d}{\lambda^{\textrm{crit}}_{d}}\Bigl)^2=0.
\label{PB}
\end{equation}
We have defined $(\lambda^{\textrm{crit}}_{t,d})^2\equiv-\hbar\omega_\textrm{rec}(\Delta^2_{t,d}+\kappa^2)/(4\Delta_{t,d})$. Self-organization occurs when $r<0$ and results from the simultaneous coupling to the two birefringent modes.

\begin{figure}[h!]
	\centering
		\includegraphics[width=\columnwidth]{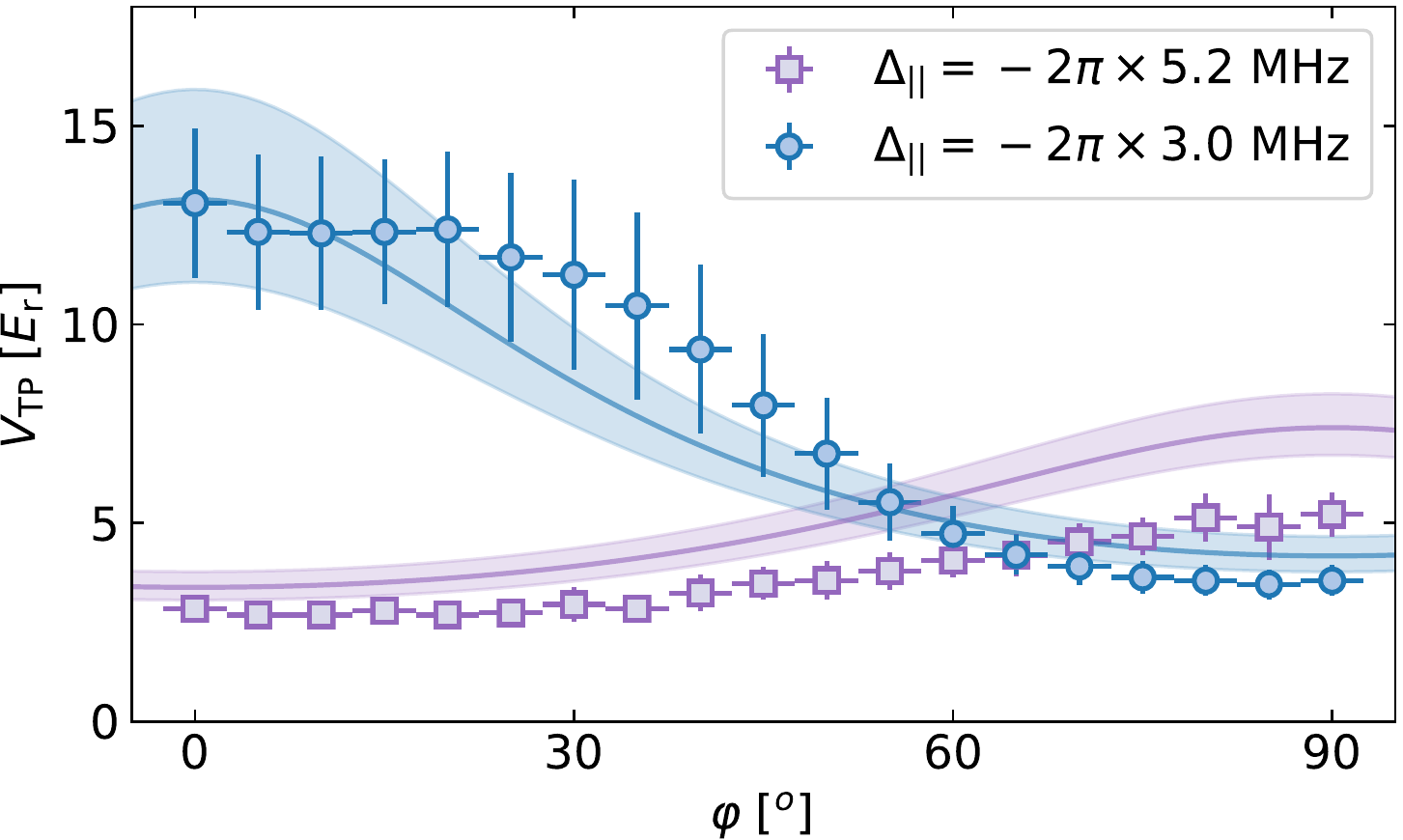}
	\caption{ \textbf{Effect of the mode competition on the phase boundary.} In two independent measurements, we set $\Delta_{\parallel}/(2\pi)=-3.0(3)$\,MHz and $\Delta_{\parallel}/(2\pi)=-5.2(3)$\,MHz and ramp up the transverse pump lattice depth from zero to $12\,\hbar\omega_\textrm{rec}$. We extract the phase boundary for different magnetic field configurations by setting a threshold on the intra-cavity photon number at $4$. Each point is an average over up to 10 repetitions. The error bar is the standard deviation extracted from the consecutive measurements of the photon threshold and includes a 10\% uncertainty on the TP lattice depth. The solid lines are the theory prediction and the shaded areas include an experimental uncertainty of 10\% on the measurement of the dispersive shift. We attribute the residual mismatch of the theory curve wih the experimental data to the variable overlap of the BEC with the transverse pump field.}
	\label{fig:competition}
\end{figure}
In particular, when $\Delta_t<0$ and $\Delta_d>0$, $(\lambda^\textrm{crit}_t)^2$ is positive but $(\lambda^\textrm{crit}_d)^2$ is negative and the critical lattice depth increases, as observed in the experiment. In a certain range of detunings, the condition $r<0$ has no real solution and self-organization is forbidden (see grey area in Fig.~\ref{fig:PDFieldAlongCavityAxis}).  The black dashed lines in Fig.~\ref{fig:PDFieldAlongCavityAxis} show the theoretical prediction for the phase boundary which is in good agreement with the experimental data. In addition, from the steady state values of $t$ and $d$ (see Eq.~S7), we can see that when a density modulation $J_x$ is formed on the BEC, both cavity modes are populated with photons. This is also reflected in our experimental data.

A transition from a two-mode to a one-mode Dicke model is observed by changing the vectorial part of the coupling independently from the scalar one. Experimentally, we align the pseudo-spin $\hat{\bf F}$ at different angles $\varphi$ in the ${\bf e}_\textrm{c}-{\bf e}_\textrm{z}$ plane. Fig.~\ref{fig:OrientingB}{\bf a-d} shows the resulting phase diagrams for $\varphi=30^\circ,\,60^\circ,\,80^\circ,\,90^\circ$. As the vectorial part of the coupling is diminished, the effect of the $\perp$-polarization mode of the cavity decreases and the system is described by a single mode Dicke model (Fig.~\ref{fig:OrientingB}{\bf d}). The measured phase diagrams agree well with the numerical simulations (see Fig.~S1). We attribute the residual feature in the $\perp$ mode in Fig.~\ref{fig:OrientingB}{\bf d} to imperfect alignment of the pump polarization with the atomic spin.

The competing character of the two modes is visible in the shift of the critical point of the SO$_\parallel$ and SO$_\perp$ phases when the vectorial part of the coupling is changed. At fixed cavity detunings $\Delta_\parallel/(2\pi)=-3.0\,$MHz and $\Delta_\parallel/(2\pi)=-5.2\,$MHz, we measure the phase boundary for different angles $\varphi$. The result of the measurement is shown in Fig.~\ref{fig:competition}. For small angles $\varphi$, when the vectorial coupling to the $\perp$-mode is large, the critical lattice depth has higher (lower) values in the SO$_\parallel$ (SO$_\perp$) phase. As the angle $\varphi$ increases, the coupling to the $\perp$ mode decreases according to $\alpha_\textrm{v}\frac{E_\textrm{p}E_0}{2\sqrt{2}}\frac{m_\textrm{F}}{2F}\,\textrm{cos}\,\varphi$. The critical point is shifted to higher or lower critical lattice depth, depending on the cavity detuning, as it is shown by the solid lines in Fig.~\ref{fig:competition}.

In conclusion we have shown that we can engineer simultaneously scalar and vectorial atom-light couplings to two orthogonal polarization modes of an optical cavity. Our experimental results can be cast into a non-degenerate two-mode Dicke model where the strength of the vectorial component of the atomic polarizability can be tuned independently from the scalar component. Measuring the change in the critical point of the self-organization phase transitions, we  demonstrated how the simultaneous presence of these interactions results in scenarios of competition. These results offer promising perspective to realize intertwined phases of spin and density degrees of freedom.

\begin{acknowledgments}
We thank Nishant Dogra and Manuele Landini for stimulating discussions. We acknowledge funding from SNF: project numbers 182650 and175329 (NAQUAS QuantERA) and NCCR QSIT, from EU Horizon2020: ERCadvanced grant TransQ (project Number 742579) and ITN grant ColOpt (project number 721465), from SBFI (QUIC, contract No. 15.0019).
\end{acknowledgments}

\beginsupplement

\subsection*{Experimental Details}

\noindent \textbf{Setup and preparation of the Bose-Einstein condensate (BEC).} We prepare an almost pure BEC of $N=3.5(3)\times 10^5$ atoms in an optical dipole trap formed by two orthogonal laser beams at a wavelength of $1064\,\text{nm}$ along the $x$-- and $y$--axes. The trapping frequencies are $(\omega_\textrm{x}, \omega_\textrm{y}, \omega_\textrm{z})=2\pi \times (120(2), 78(1), 193(2))\,\mathrm{Hz}$. The trap position coincides with the center of two TEM$_{00}$ optical modes ($\parallel,\,\perp$) of a high-finesse cavity \cite{Leonard2017}. The atom number is extracted by measuring the magnitude of the dispersive shift $NU_0/(2\hbar)=170(10)\,$kHz of the cavity resonance in the presence of the BEC. The intra-cavity lattice depth per photon $U_0$ is calculated from the geometry of the cavity and taking into account the D$_1$ and D$_2$ atomic lines. The cavity has a birefringence of $\delta=2\pi\times3.89(1)\,$MHz between the horizontally ($\perp$) and vertically ($\parallel$) polarized modes. We can adjust the resonance frequency of the cavity mode with piezoelectric elements that are included in the mount of each cavity mirror. The frequency is actively stabilized with the help of an additional laser beam at $\sim830\,\text{nm}$. The corresponding residual intra-cavity lattice potential of $\sim0.1\,E_\text{rec}$ is negligible with respect to the critical lattice depth for the self-organization phase transition and incommensurate with the cavity mode.

\noindent \textbf{Lattice and photon number calibrations.} In order to calibrate the lattice depths of the transverse pump and each cavity field we perform Raman-Nath diffraction on the atomic cloud. The intra-cavity photon number calibration can be calculated from the lattice depth per photon $U_0$. We extract detection efficiencies of $1.94(1)\%$ and $1.0(1)\%$ for detecting with single-photon counting modules an intra-cavity photon in the mode $\parallel$ and $\perp$, respectively.

\noindent \textbf{Detection of the cavity output polarization.} The intra-cavity light field leaking from the mirror is split on a polarizing beam splitter (PBS) that separates the reflected ($\alpha_\textrm{R}$) and transmitted ($\alpha_\textrm{T}$) part of the light field onto single photon detectors. Due to imperfect alignment of the cavity modes $\perp,\,\parallel$ with the PBS axis we calibrate $\alpha_{\parallel}=\sqrt{n_\parallel}$ and $\alpha_{\perp}=\sqrt{n_\perp}$ from $\alpha_\textrm{T}$ and  $\alpha_\textrm{R}$.  The results are plotted in Fig.~2 and Fig.~3.

\noindent \textbf{Magnetic field calibration and manipulation.}
To calibrate the direction of the magnetic field we perform radio-frequency spectroscopy on the atomic cloud. Using three sets of magnetic coils aligned orthogonal to each other, we can identify the orientation of the residual magnetic field in the laboratory at the position of the atoms and compensate it. Before spontaneous demagnetization of the cloud occurs due to magnetic noise, we can reduce the magnetic field at the BEC position to values as low as $\sim14\,$mG, which corresponds to a Zeeman shift of about $\sim10\,$kHz for atoms in the $F=1$ manifold. We use these three pairs of coils to orient the magnetic field in space. Each pair of coils allows to maximally generate a field of  $\sim3\,$G at the BEC position. All data shown in the main text are taken with a large offset field $B\geq4\,$G, creating a Zeeman level splitting larger than $\Delta_{\perp}$. In this way, collective cavity-pump Raman transitions between different Zeeman sub-levels are suppressed. For $\varphi=90^\circ$ we use an additional pair of magnetic offset coils.

\subsection*{Theory}
\noindent \textbf{Quantum--optical Hamiltonian.}
We now consider the effect of the pseudo-spin  $\hat{\bf F}$ of the atoms on the self-organization phase transition in our setup. Here we show how the geometry of our pump polarization and cavity birefringence generates the hamiltonian in the main text. The role of the atomic spin on the self-organization phase transition has been recently studied \cite{Landini2018}. In our theoretical description, we refer to the system depicted in Fig.~1, where the (pseudo-) spin is aligned in the ${\bf e_\textrm{z}}-{\bf e_\textrm{c}}$ plane. The two orthogonal birefringent modes of the cavity are labelled $\perp$ for the horizontal mode and $\parallel$ for the vertical mode. The total electric field operator ${\bf \hat{E}}$ is the sum of the coherent pump field ${\bf {E}_\textrm{p}}$ and the quantized $\perp$ (${\bf E}_{\perp}$) and $\parallel$ (${\bf E}_{\parallel}$) fields of the cavity
\begin{align}
{\bf \hat{E}}={\bf E}_\textrm{p}+{\bf E}_\parallel\hat{a}+{\bf E}_\perp\hat{b}.
\end{align}
Here, we assume the pump laser to be a classical field while the $\parallel$ and $\perp$ modes are quantized fields with associated annihilation operators $\hat{a}$ and $\hat{b}$.

The atom-light interaction hamiltonian can be written in the following form \cite{Kien2013}
\begin{align}
\hat{H}_\textrm{int}=-{\bf \hat{E}}^\dagger\underline{\alpha}{\bf \hat{E}}=-\alpha_\textrm{s}{\bf \hat{E}}^\dagger\cdot{\bf \hat{E}}+i\alpha_\textrm{v}\Bigl({\bf \hat{E}}^\dagger\wedge{\bf \hat{E}}\Bigl)\cdot\frac{{\bf \hat{F}}}{2F}\,,
\end{align}
where $\underline{\alpha}$ is the polarizability tensor of the atom composed of a scalar ($\alpha_\textrm{s}$) and a vectorial part ($\alpha_\textrm{v}$). The electric fields ${\bf E}_\textrm{p}=\frac{E_\textrm{p}}{2}\,\textrm{cos}\,(k_\textrm{p}y)\,{\bf e}_\textrm{p}$, ${\bf E}_{\parallel}=E_{0}\,\textrm{cos}\,(\bf{k}_\textrm{c}\cdot\bf{r})\,{\bf e}_{\parallel}$ and ${\bf E}_{\perp}=E_{0}\,\textrm{cos}\,(\bf{k}_\textrm{c}\cdot\bf{r})\,{\bf e}_{\perp}$  are linearly polarized, with ${\bf e}_\textrm{p}={\bf e}_{\parallel}={\bf e}_\textrm{z}$, ${\bf e}_{\perp}=\frac{\sqrt{3}}{2}{\bf e_\textrm{y}}-\frac{1}{2}{\bf e_\textrm{x}}$ and mode functions $\textrm{cos}\,(\bf{k}_\textrm{c}\cdot\bf{r})$ and $\textrm{cos}\,({k}_\textrm{p}y)$. ${\bf e_\textrm{x}},\,{\bf e_\textrm{y}},\,{\bf e_\textrm{z}}$ are the unit vectors pointing along $x$, $y$ and $z$ (see Fig.~1). For this choice of polarizations, the scalar part of the hamiltonian has the form
\begin{align}
\hat{H}_\textrm{s}&=-\alpha_\textrm{s}{\bf \hat{E}}^\dagger\cdot{\bf \hat{E}}=-\alpha_\textrm{s}{\bf E}^2_\textrm{p}-\alpha_\textrm{s}{\bf E}^2_{\parallel}\hat{a}^\dagger\hat{a}\notag\\
&-\alpha_\textrm{s}{\bf E}^2_{\perp}\hat{b}^\dagger\hat{b}-\alpha_\textrm{s}{\bf E}_\textrm{p}\cdot{\bf E}_{\parallel}\Bigl(\hat{a}^\dagger+\hat{a}\Bigl),
\label{ScalarEnergy}
\end{align}
which is the standard form describing self-organization \cite{Baumann2010}. The vectorial part can be evaluated carrying out the vector products
\begin{align}
\hat{H}_\textrm{v}&=i\alpha_\textrm{v}\Bigl({\bf E}_\textrm{p}+{\bf E}_{\parallel}\hat{a}^\dagger+{\bf E}_{\perp}\hat{b}^\dagger\Bigl)\wedge\Bigl({\bf E}_\textrm{p}+{\bf E}_{\parallel}\hat{a}+{\bf E}_\perp \hat{b}\Bigl)\cdot\frac{\hat{\bf F}}{2F}\notag\\
&=i\alpha_\textrm{v}\Bigl[\bigl({\bf E}_\textrm{p}\wedge{\bf E}_{\perp}\bigl)\,\hat{b}+\bigl({\bf E}_{\parallel}\wedge{\bf E}_{\perp}\bigl)\,\hat{a}^\dagger\hat{b}+\bigl({\bf E}_{\perp}\wedge{\bf E}_\textrm{p}\bigl)\,\hat{b}^\dagger\notag\\
&+\bigl({\bf E}_{\perp}\wedge{\bf E}_{\parallel}\bigl)\,\hat{b}^\dagger\hat{a}\Bigl]\cdot\frac{\hat{\bf F}}{2F}\notag\\
&=i\alpha_\textrm{v}\Bigl[E_{0}^2\Bigl(\hat{a}^\dagger\hat{b}-\hat{b}^\dagger\hat{a}\Bigl)\textrm{cos}^2({\bf k}_\textrm{c}\cdot{\bf r})\notag\\
&+E_\textrm{p}E_{0}(\hat{b}-\hat{b}^\dagger)\,\textrm{cos}\,({\bf k}_\textrm{c}\cdot{\bf r})\,\textrm{cos}\,(k_\textrm{p}y)\Bigl]\Bigl({\bf e}_\textrm{c}\cdot\frac{\hat{\bf F}}{2F}\Bigl),
\label{VectorialEnergy}
\end{align}
where ${\bf e}_\textrm{c}$ is the unit vector along the direction of the cavity axis (see Fig.~1).

The first term in the result of Eq.~\ref{VectorialEnergy} describes a direct coupling between the modes $\perp$ and $\parallel$, whereas the second term is a coupling between the transverse pump and the out of phase quadrature of the $\hat{b}$ mode. Since the vectorial part of the hamiltonian is proportional to the product $\Bigl({\bf e}_\textrm{c}\cdot\hat{\bf F}\Bigl)$ between the pseudo-spin $\hat{\bf F}$ and the cavity unit vector ${\bf e}_\textrm{c}$, its contribution vanishes if they are orthogonal to each other.\\
In the rotating frame of the pump, we can write the many-body hamiltonian of the BEC-cavity system including the energy cost of intra-cavity photons, the kinetic energy of the atoms and the contributions from Eq.~\ref{ScalarEnergy}, Eq.~\ref{VectorialEnergy} as,
\begin{align}
\hat{H}_\textrm{BEC}&=-\Delta_{\parallel}\hat{a}^\dagger\hat{a}-\Delta_{\perp}\hat{b}^\dagger\hat{b}\notag\\
&+\int\hat{\Psi}^\dagger({\bf r})\Bigl(\frac{\hat{p}^2}{2m}+\hat{H}_\textrm{s}+\hat{H}_\textrm{v}\Bigl)\hat{\Psi}({\bf r})d{\bf r}.
\end{align}
Following the procedure reported in the supplementary material of \cite{Leonard2017}, we restrict the ansatz of the atomic wavefunction and derive an effective hamiltonian describing the BEC-cavity system. The resulting hamiltonian reads as
\begin{align}
\hat{H}&=-\hbar\Delta_{\parallel}\hat{a}^\dagger\hat{a}-\hbar\Delta_{\perp}\hat{b}^\dagger\hat{b}+\hbar\omega_-\hat{c}_{-}^\dagger\hat{c}_{-}+\hbar\omega_+\hat{c}_{+}^\dagger\hat{c}_{+}\notag\\
&+\frac{1}{2\sqrt{2}}\alpha_\textrm{s}E_\textrm{p}E_\textrm{0}\Bigl(\hat{a}^\dagger+\hat{a}\Bigl)\Bigl(\hat{c}_-^\dagger\hat{c}_0+\hat{c}_+^\dagger\hat{c}_0+\textrm{h.c.}\Bigl)\notag\\
&+i\alpha_\textrm{v}\Bigl[\frac{1}{2\sqrt{2}}E_\textrm{p}E_{0}\Bigl(\hat{b}^\dagger-\hat{b}\Bigl)\Bigl(\hat{c}_-^\dagger\hat{c}_0+\hat{c}_+^\dagger\hat{c}_0+\textrm{h.c.}\Bigl)\notag\\
&+\frac{1}{2}E_\textrm{0}^2\Bigl(\hat{a}^\dagger\hat{b}-\hat{b}^\dagger\hat{a}\Bigl)\hat{c}_0^\dagger\hat{c}_0\Bigl]\Bigl({\bf e}_\textrm{c}\cdot\frac{\hat{\bf F}}{2F}\Bigl),
\label{CompleteSpinHamiltonian}
\end{align}
where $\Delta_{{\parallel/\perp}}$ include the dispersive shift of the cavity resonance due to the atoms. $\hat{c}_0,\,\hat{c}_-,\,\hat{c}_+$ ($\hat{c}^\dagger_0,\,\hat{c}_-^\dagger,\,\hat{c}_+^\dagger$) are the annihilation (creation) operators of an atom in the zero, ${\bf k}_+={\bf k}_\textrm{p}+{\bf k}_\textrm{c}$ and ${\bf k}_-={\bf k}_\textrm{p}-{\bf k}_\textrm{c}$ momentum states, with associated energy zero, $\omega_+=(\hbar{\bf k}_+)^2/2m$ and $\omega_-=(\hbar{\bf k}_-)^2/2m$, respectively. We define the amplitude of the transverse pump lattice as $V_\textrm{TP}=-\alpha_\textrm{s}E^2_\textrm{p}/4$ and the amplitude of the cavity lattice as $V_\textrm{c}=-\alpha_\textrm{s}E^2_\textrm{0}$.  We neglect the dissipation of the cavity, the contribution of the transverse pump lattice $\alpha_\textrm{s}{ E}^2_\textrm{p}$, atomic collisions and the effect of the trapping potential.  Since we work at high magnetic fields, we can write ${\bf e}_\textrm{c}\cdot\frac{\hat{\bf F}}{2F}=\frac{m_\textrm{F}}{2F}\,\textrm{cos}\,\varphi$, where the angle $\varphi$ is shown in Fig.~3.
Since the operator $\hat{c}_-$ creates atoms in the lowest momentum state, we can obtain the low-energy theory of the system neglecting in Eq.~\ref{CompleteSpinHamiltonian} the terms involving the operator $\hat{c}_+$. Setting $\hat{c}\equiv\hat{c}_-$ we obtain the hamiltonian Eq.~2 in the main text. \\
\noindent \textbf{Mean-field solution of the two-mode Dicke model.}
From the hamiltonian Eq.~4 in the main text, we can derive a steady-state solution for the spin and the photon operators in a mean-field limit. Given the different timescales for the atomic and the light field evolution, we can adiabatically eliminate the latter by considering its steady-state values
\begin{equation}
\begin{split}
\hat{t} &= \frac{ \lambda_{t}}{\sqrt{N}}\frac{\hat{J}_x}{\Delta_{t} + i \kappa}, \\
\hat{d} &= \frac{ \lambda_{d}}{\sqrt{N}}\frac{\hat{J}_x}{\Delta_{d} + i \kappa}.
\label{eq:photonsteadystate}
\end{split}
\end{equation}
For the spin operators, we derive the following system of equations
\begin{equation}
\begin{cases}
\dot {\hat{J}}_\textrm{x}  = -\omega_\textrm{rec} \hat{J}_\textrm{y}
\\
\dot {\hat{J}}_\textrm{y}  = \omega_\textrm{rec} \hat{J}_\textrm{x} - \frac{2\eta}{N} J_\textrm{z}
\\
\dot J_\textrm{z}  =  \frac{2\eta}{N}J_\textrm{y}
\end{cases}
\end{equation}
where we defined
\begin{equation}
\eta = \frac{4 \Delta_t\lambda_\textrm{t}^2}{\Delta_\textrm{t}^2+\kappa^2}+\frac{ 4\Delta_\textrm{d}\lambda_\textrm{d}^2}{\Delta_\textrm{d}^2+\kappa^2}.
\end{equation}
To find the steady state value of $J_\textrm{x}$, we take the steady state solution $J_\textrm{y} = 0$ and the further substitution $J_\textrm{z}  =  -\sqrt{\frac{N^2}{4} - J_\textrm{x}^2}$. In addition to the trivial solution $J_\textrm{x} = 0$, two roots satisfy the quadratic equation
\begin{equation}
J_\textrm{x}^2 = \frac{N^2}{4} \left( 1-\frac{\omega_\textrm{rec} ^2}{ \eta^2} \right)
\label{eq:Jxsteadystate}
\end{equation}
above the self-organisation phase threshold, which is then given by the condition
\begin{equation}
1 - \frac{4 \Delta_\textrm{t}\lambda_\textrm{t}^2}{\omega_\textrm{rec}(\Delta_\textrm{t}^2+\kappa^2)}-\frac{ 4\Delta_\textrm{d}\lambda_\textrm{d}^2}{\omega_\textrm{rec}(\Delta_\textrm{d}^2+\kappa^2)} = 0.
\end{equation}

From Eq.~\ref{eq:Jxsteadystate} and Eq.~\ref{eq:photonsteadystate} we derive the photon numbers $\langle\hat{a}^\dagger\hat{a}\rangle$ and $\langle\hat{b}^\dagger\hat{b}\rangle$, after applying the inverse unitary transformations Eq.~3.
\begin{figure}[h!]
	\centering
		\includegraphics[width=0.80\columnwidth]{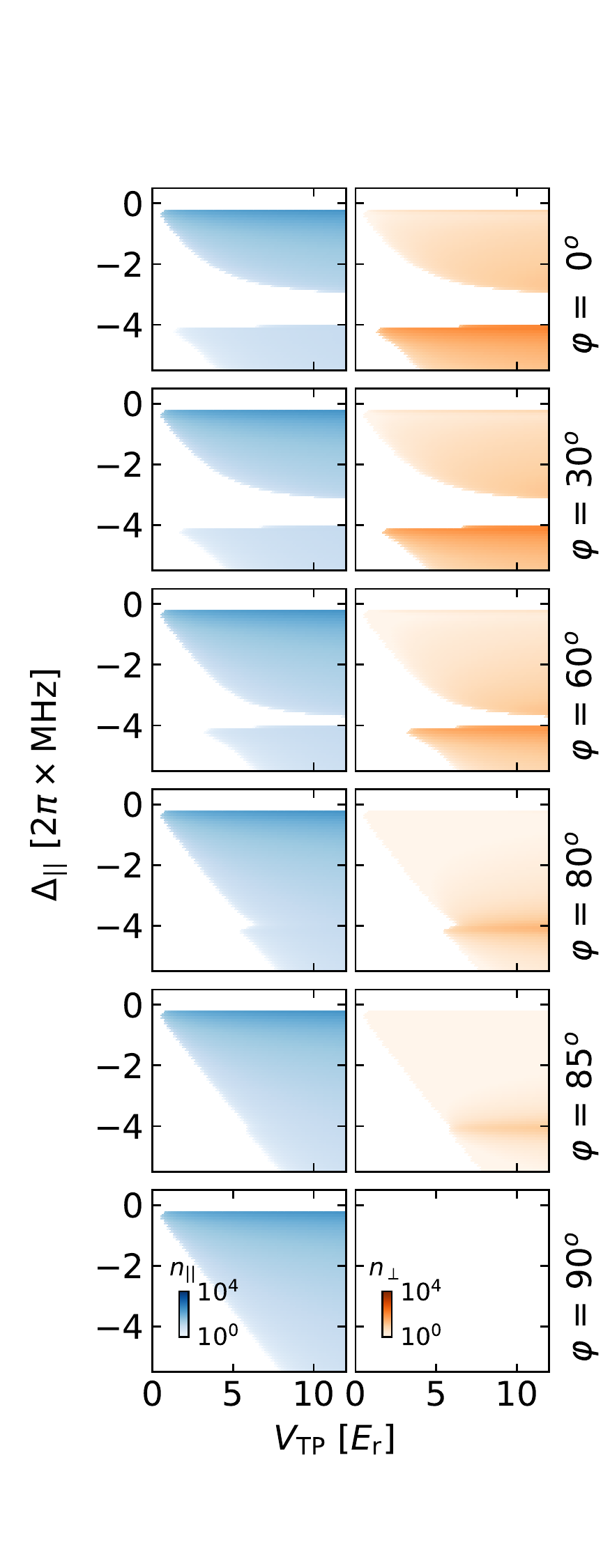}
	\caption{\textbf{Mean-field solution of the phase diagram.} The photon levels $n_\parallel$ and $n_\perp$ are shown as a function of the transverse pump lattice depth ($V_\textrm{TP}$) and cavity detuning ($\Delta_{\parallel}$) for different values of the angle $\varphi$. The transverse pump and cavity detuning ranges are chosen equivalent to the data shown in Fig.~3.  }
	\label{fig:Numerics}
\end{figure}
The result of the numerical calculation for the phase diagram for different angles $\varphi$ is reported in Fig.~\ref{fig:Numerics}.

\bibliographystyle{apsrev4-1}

%

\appendix
\end{document}